\documentclass[aps,prl,twocolumn,showpacs,groupedaddress]{revtex4}
\usepackage{graphicx}

\begin{document}               

\def\be{\begin{equation}}
\def\ee{\end{equation}}
\def\bd{\begin{displaymath}}
\def\ed{\end{displaymath}}
\def\ba{\begin{eqnarray}}
\def\ea{\end{eqnarray}}
\def\lr{\leftrightarrow}
\def\s{\protect{f_0}}

\title{Closed-flavor $\pi J/\psi$ and   $\pi \Upsilon $ Cross Sections 
at Low Energies\\ from Dipion Decays}
\author{T.Barnes}
\email{tbarnes@utk.edu}
\affiliation{Physics Division, 
Oak Ridge National Laboratory, Oak Ridge, TN 37831, USA,\\
Department of Physics and Astronomy, 
University of Tennessee,\\
Knoxville, TN
37996, USA}
\author{N.I.Kochelev}
\email{kochelev@thsun1.jinr.ru}
\affiliation{
Bogoliubov Laboratory of Theoretical Physics,
Joint Institute for Nuclear Research,\\ 
141980 Dubna, Moscow Region, Russia,
Institute of Physics and Technology\\
480082 Almaty, Kazakhstan
}
\date{\today}

\begin{abstract}
The scale of low energy
$c\bar c$ and $b\bar b$ cross sections on light hadrons is
of great importance to searches for the quark gluon
plasma using the heavy-quarkonium suppression
signature. 
Unfortunately, little is known about these 
near-threshold cross sections at present, and
recent theoretical estimates span many orders of magnitude. 
Here we use 
experimental data on the four observed
closed-flavor heavy quarkonium hadronic decays
$\psi{\, '}\to \pi \pi J/\psi $,
$\Upsilon' \to \pi \pi \Upsilon$,
$\Upsilon'' \to \pi \pi \Upsilon $
and
$\Upsilon'' \to \pi \pi \Upsilon' $,
combined with simple models of the transition amplitudes,
to estimate the pion scattering cross sections 
of $c\bar c$ and $b\bar b$ mesons
near threshold. Specifically we consider the closed-flavor reactions
$\pi J/\psi \to  \pi \psi{\, '}$,
$\pi \Upsilon \to \pi \Upsilon' $,
$\pi \Upsilon \to \pi \Upsilon'' $
and
$\pi \Upsilon' \to \pi \Upsilon'' $ and their 
time-reversed analogues.
Our results
may be useful in constraining theoretical models
of the strong interactions of heavy quarkonia,
and can be
systematically improved through future detailed studies of 
dipion decays, notably
$\psi{\, '}\to \pi\pi J/\psi$ and
$\Upsilon'' \to \pi\pi\Upsilon$.
\end{abstract}


\maketitle

\section{Introduction}

One signature proposed for the identification of a 
quark-gluon plasma (QGP) is a suppressed rate of formation
of charmonium bound states, due to the screening 
effect of the plasma on the linear
confining potential that normally binds a $c\bar c$ pair 
\cite{Matsui:1986dk}. 
In the presence of this screening, 
$c\bar c$ pairs produced within the plasma presumably 
separate into open charm mesons, so that the formation of a QGP 
would be signaled by a decrease in the charmonium production cross section.

This attractively simple picture becomes more complicated if the dissociation
cross sections of charmonia on light hadrons are not small. In this case 
the charmonia produced during a heavy-ion collision may rescatter
into open-charm final states as the result of interactions with 
the ``comoving" light hadrons produced in the collision.

Due to the importance of these cross sections for QGP searches,
many calculations of near-threshold scattering cross sections 
of light hadrons on charmonia have recently 
been reported. The methods applied include
a high energy color-dipole scattering model \cite{Kharzeev:1994pz}, 
quark models \cite{Martins:1994hd,Wong:1999zb,Barnes:2003dg},
meson exchange models \cite{Matinian:1998cb,Haglin:1999xs,Navarra:2001pz}, 
and most recently 
QCD sum rules \cite{Navarra:2001jy,Duraes:2002ux}.  

Under certain simplifying assumptions one may relate 
experimentally known
hadron decays 
to heavy quarkonium scattering cross sections.
Here we consider the 
dipion decays of
heavy quarkonia below open-flavor thresholds, which
have been observed in four cases,
$\psi{\, '} \to \pi \pi J/\psi\, $
\cite{Armstrong:pg,Bai:1999mj,Ambrogiani:2000vj},
$\Upsilon' \to  \pi \pi\Upsilon$
\cite{Besson:1984ha,Bowcock:1986um,Brock:pj,Butler:1993rq}
$\Upsilon'' \to \pi \pi\Upsilon$
\cite{Bowcock:1986um,Brock:pj,Heintz:cv,Wu:qc,Butler:1993rq}
and
$\Upsilon'' \to \pi \pi\Upsilon'$
\cite{Bowcock:1986um,Brock:pj,Heintz:cv,Wu:qc,Butler:1993rq}.
These transitions are quite weak, with partial widths of 
only {\it ca.}
100~keV for $c\bar c$ and 1$-$10~keV for $b\bar b$. 
They have been observed only because the total widths of the initial 
heavy quarkonia, which lie below their open-flavor thresholds,
are also very small.

We can use these three-body partial
widths to estimate pion scattering cross sections of the 
corresponding heavy quarkonia, since these processes 
are described by
the same invariant amplitudes.
The 
pion scattering cross sections are given by
\be
\sigma_{\pi_1 {\cal M}_a \to \pi_2 {\cal M}_b } 
= 
{1\over 64\pi s}\, {1\over p_i^{\, 2}}  
\int_{t_1}^{t_2} \! dt\; \langle\, |{\cal A}\, |^2  \rangle  
\ee
where as usual
$
s 
= m_{\pi_1 {\cal M}_a }^2
= m_{\pi_2 {\cal M}_b }^2
$, 
$t = (p_{\pi_1} - p_{\pi_2})^2$, 
the limits are 
$t_{2\atop 1} = -2[ E_{\pi_1}E_{\pi_2}\mp p_ip_f - m_{\pi}^2 ]$,
and frame-dependent quantities 
such as $ E_{\pi_1}$ and $p_i = |\vec p_i |$ are understood to be 
evaluated in the c.m. frame.

The corresponding 
dipion three-body partial width is 
\be
\Gamma_{{\cal M}_b \to \pi_1 \pi_2 {\cal M}_a }
=
{1\over (2\pi)^3}\,
{1\over 32 {M_b}^3}
\int \!\!\!
\int \! 
ds\, dt \;
\langle\, |{\cal A}\, |^2  \rangle  
\ee
where $s$ and $t$ (after crossing) become 
$s = m_{\pi_2 {\cal M}_a }^2$ and
$t = m_{\pi_1 \pi_2}^2$. 
Our Eqs.(1,2) correspond to  
Eqs.(37.30,21) 
of the 2002 PDG \cite{Hagiwara:fs}.

Our convention is that ${\cal M}_b$ is higher in mass
than ${\cal M}_a$, so the dipion decay 
${\cal M}_b \to \pi \pi {\cal M}_a$ 
is energetically
allowed but
${\cal M}_a \to \pi \pi {\cal M}_b$ 
is not.
The initial and final mesons 
in the cross section formula
Eq.(1)
may be
$({\cal M}_a,{\cal M}_b)$ 
as shown, or they may be
transposed to
$({\cal M}_b,{\cal M}_a)$,
since the same invariant amplitude
${\cal A}$ is involved. 
As an example, the decay
$\psi{\, '} \to \pi^+ \pi^- J/\psi $ 
is related to both
$\pi^+ J/\psi \to \pi^+ \psi{\, '} $ 
and
$\pi^+ \psi{\, '} \to \pi^+ J/\psi$. 

Since the squared invariant amplitude 
$\langle\, |{\cal A}\, |^2  \rangle$
is sampled in different kinematic regions by 
the decay and the cross sections
(see Fig.1), assumptions regarding the form of
$\langle\, |{\cal A}\, |^2  \rangle$ are required to relate
these various processes. In the following we shall obtain
results for the cross sections given three simple models
of the invariant amplitudes. 

\begin{figure}
\includegraphics[width=0.4\textwidth,angle=270]{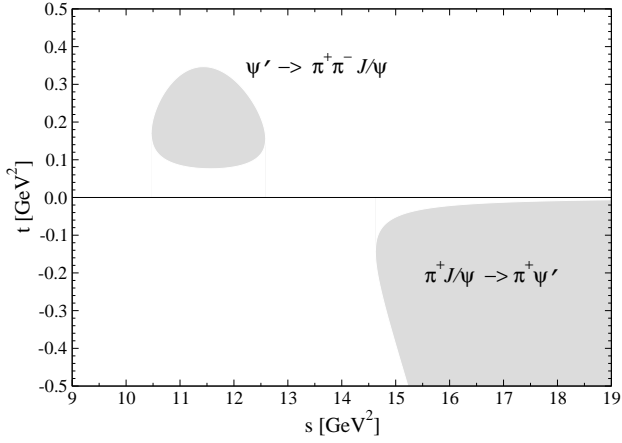}
\vskip -0.3cm
\caption{Dalitz plot of kinematically allowed regions for the
decay
$\psi{\, '}\to \pi^+ \pi^- J/\psi $ and the
reactions
$
\pi^+ J/\psi 
\to 
\pi^+ \psi{\, '}
$ 
and 
$
\pi^+ \psi{\, '}
\to 
\pi^+ J/\psi 
$
. 
} 
\label{fig:bk_fig1}
\end{figure}

\begin{figure}[ht]
\includegraphics[angle=270,width=0.5\textwidth]{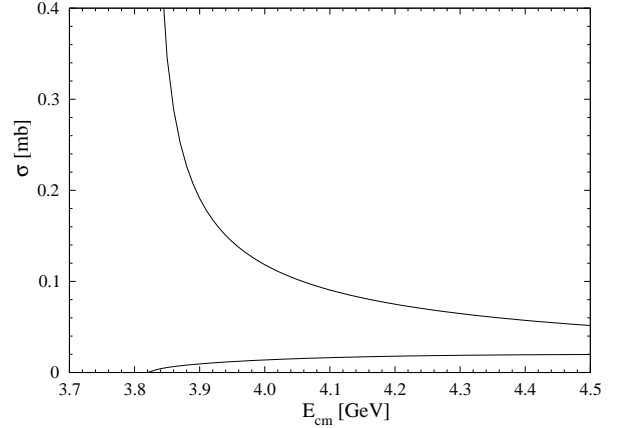}
\vskip -0.5cm
\caption{
$\sigma_{\pi^+ J/\psi \to \pi^+ \psi{\, '}}$ (lower) 
and
$\sigma_{\pi^+ \psi{\, '} \to \pi^+ J/\psi}$ (upper)
estimated from 
$\Gamma_{\psi{\, '} \to  \pi^+ \pi^- J/\psi\, }$
assuming constant amplitudes.
} 
\label{fig:bk_fig2}
\end{figure}

\vskip -0.5cm
\begin{figure}[ht]
\includegraphics[angle=270,width=0.5\textwidth]{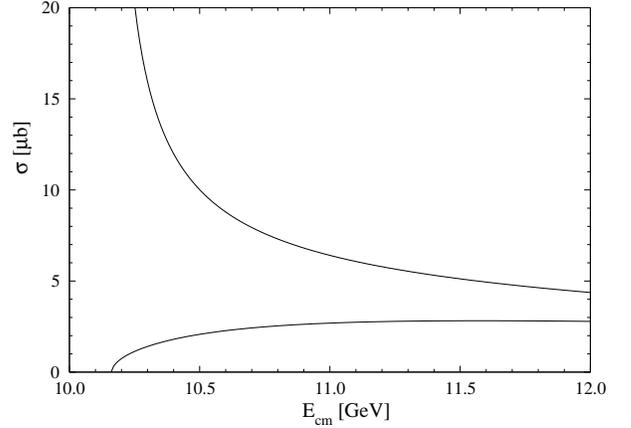}
\vskip -0.5cm
\caption{
$\sigma_{\pi^+ \Upsilon \to \pi^+ \Upsilon'}$ (lower) 
and
$\sigma_{\pi^+ \Upsilon' \to \pi^+ \Upsilon}$  
(upper) estimated from 
$\Gamma_{\Upsilon' \to \pi^+ \pi^- \Upsilon }$
assuming constant amplitudes.} 
\label{fig:bk_fig3}
\end{figure}

\section{Constant ${\cal A}$ approximation}

As a first approximation  
we neglect any dependence of $\langle\, |{\cal A}\, |^2  \rangle$
on kinematics, and simply treat it as a constant. In this case
Eqs.(1,2) imply a simple relation between the pion cross
section and the dipion partial decay width, 
\be
\sigma_{ \pi {\cal M}_a \to \pi {\cal M}_b } 
= 
\Gamma_{ {\cal M}_b \to \pi\pi {\cal M}_a } 
\cdot
{16\pi^2 
M_b^3\over A_D}  
\; 
{p_f\over p_i}
\,
s^{-1}
\equiv 
c_0
\,
{p_f\over p_i}
\,
s^{-1}
\ee  
where $A_D$ is the area of the 
${\cal M}_b \to \pi\pi {\cal M}_a$
dipion decay 
Dalitz plot
\be
A_D 
=
\int \!\!\!
\int \!
dm_{\pi_2 {\cal M}_a }^2 
dm_{\pi_1 \pi_2 }^2 
=
\int \!\!\!
\int \!
ds \, dt
\ee 
and $p_i = p_a $ and $p_f = p_b$ 
are the three-momenta of the initial and final
pions (or heavy mesons) in the reaction
$\pi {\cal M}_a \to \pi {\cal M}_b$ in the $c.m.$ frame. 
The 
Dalitz plot for the 
decay 
$\psi{\, '} \to \pi^+ \pi^- J/\psi $ 
and the related
reactions 
$\pi^+ J/\psi \to \pi^+ \psi{\, '} $ 
and
$
\pi^+ \psi{\, '} 
\to 
\pi^+ J/\psi 
$
are shown as examples in Fig.1.
The Dalitz plot areas $\{ A_D\} $ and dipion widths used in
this work are given in Table~1; the masses assumed are
$m_{\pi^+}  = 0.1396$~GeV,
$M_{J/\psi} = 3.097$~GeV,
$M_{\psi{\, '}}  = 3.686$~GeV,
$M_{\Upsilon}  = 9.460$~GeV,
$M_{\Upsilon'}  =  10.023$~GeV and
$M_{\Upsilon''}  = 10.355$~GeV.
 
\begin{table}[h]
\begin{center}
\begin{tabular}{|c|c|c|c|} 
\hline
transition 
& 
\
$A_D$[GeV$^4$] 
\
& 
\
$\Gamma_{\pi^+\pi^-}$[keV] 
\
& 
\
$c_0$[mb GeV$^2$] 
\
\\ \hline \hline
\
$\psi{\, '}\to\pi^+\pi^- J/\psi$ 
\
& $0.436$ & $91.5\pm 9.0$ 
& $0.65\pm 0.06 $
\\ \hline
$\Upsilon'\phantom{'}\to\pi^+\pi^-\Upsilon\phantom{'}$ 
& $1.023$ & $8.3\pm 1.3$ 
& $0.50\pm 0.08$
\\ \hline
$\Upsilon''\to\pi^+\pi^-\Upsilon\phantom{'}$ 
& $6.679$ & $1.2\pm 0.2$ 
& $0.012\pm 0.001$ 
\\ \hline
$\Upsilon''\to\pi^+\pi^-\Upsilon'$ & $0.027$ & $0.7\pm 0.2$ 
& $1.9\pm 0.5 $
\\ \hline
\end{tabular} 
\caption{Experimental dipion-decay Dalitz plot areas, transition rates
and cross section coefficients.} 
\end{center}
\end{table}

The cross sections for the reactions
$\pi^+ J/\psi \to \pi^+ \psi{\, '}$ 
and
$\pi^+ \psi{\, '} \to \pi^+ J/\psi$
in the 
constant amplitude approximation are shown in Fig.2.
Evidently the scales of these cross sections 
a few hundred MeV above threshold are 
{\it ca.} 20~{$\mu$b} 
for the endothermic process
$\pi^+ J/\psi \to \pi^+ \psi{\, '}$
and {\it ca.} 0.1~{mb} for its crossed exothermic
partner 
$\pi^+ \psi{\, '} \to \pi^+ J/\psi$. This method applied to pion 
cross sections in the upsilon family
yields cross sections of
{\it ca.} 2~{$\mu$b} for
$\pi^+ \Upsilon \to \pi^+ \Upsilon'$
and 
{\it ca.} 10~{$\mu$b} for
$\pi^+ \Upsilon' \to \pi^+ \Upsilon$
in the analogous kinematic regime (Fig.3).

\section{Improved estimates}

Although the constant amplitude results of the previous section 
are of interest as order-of-magnitude
estimates, it is known experimentally that the invariant amplitudes for 
these dipion decays show a strong and rather complicated dependence
on the $\pi\pi$ invariant mass. 
In $\psi{\, '} \to\pi^+\pi^- J/\psi $ a strong
suppression of the $\pi\pi$ system at low invariant mass is evident
(see for example our Fig.4, or Fig.6 of Bai {\it et al.}\cite{Bai:1999mj}). 
One might have instead expected a near-threshold
enhancement, due to the attractive FSI of an S-wave I=0 $\pi\pi$ system. 
In contrast, the decay $\Upsilon'' \to \pi\pi \Upsilon$
has a complicated double-bump structure, 
which {\it does} show an enhancement at low $\pi\pi$ invariant mass,
as well as a high invariant mass 
enhancement as seen in 
$\psi{\, '} \to \pi\pi J/\psi$.
This difference between the observed $\pi\pi$ distributions 
suggests that the 
$\pi\pi$ production amplitude 
depends significantly on the heavy-quarkonium
source,
and is not accurately described by a universal 
$\pi\pi$ amplitude alone. 

Various theoretical models
of these dipion decay amplitudes have appeared in the
literature. They may be divided into three broad
categories, 
according to the mechanism that is assumed to give rise
to the strong $m_{\pi\pi}$-dependence;
1) gluon radiation 
models 
\cite{Gottfried:1977gp,Yan:1980uh,Chen:1997zz},
2) scalar anomaly models \cite{Voloshin:yb,Voloshin:1980zf,Novikov:fa}, and
3) scalar meson exchange models \cite{Sorge:1997bg,Komada:2000ez,Lahde:2001zd}.
The first two categories predict rather similar $m_{\pi\pi}$ 
dependences, so we will consider them together.
There is also a suggestion that chiral symmetry, combined
with certain simplifying asumptions for amplitudes, can
explain the observed $m_{\pi\pi}$ dependence in 
$\psi{\, '} \to \pi\pi J/\psi$ \cite{Brown:1975dz}; this model
suggests an $m_{\pi\pi}$ dependence similar to the
first two categories. 
Finally, we note that several references have considered the
decay 
$\Upsilon'' \to \pi\pi\Upsilon$ as a special case, since it has a
``double-humped" dipion distribution that is not seen in other
decays \cite{Moxhay:1988ri,Kiselev:1994ex}.

Gluon radiation and scalar anomaly models assume that the 
important low-energy $m_{\pi\pi}$ dependence is determined by
aspects of a purely gluonic intermediate state. In gluon radiation
models this strong energy dependence arises from a multipole expansion
of the gluon emission amplitude, 
whereas in the scalar anomaly models it is the momentum dependence
encountered in coupling the gluonic state to the $\pi\pi$ final state.
Clearly it will be difficult to distinguish these two possibilities,
although a detailed comparison of their predictions with experiment 
appears to favor the scalar anomaly model \cite{Bai:1999mj}.

In the scalar anomaly model one can relate the $\pi\pi$ 
production amplitude to the matrix element of the simplest
scalar gluon operator between the vacuum and a $\pi\pi$ state, 
$\langle \pi\pi | G_{\mu\nu}^a G_{\mu\nu}^a | 0 \rangle$. 
This matrix element can be determined because 
the operator $G_{\mu\nu}^a G_{\mu\nu}^a$ is proportional to the
triangle anomaly in the trace of the energy-momentum tensor, which in 
a low energy pion effective lagrangian is quadratic in the pion field.
The matrix element
of this operator gives a near-threshold dependence of
${\cal A} \propto m_{\pi\pi}^2$, which is ${\cal A} \propto t$ in our 
kinematics.

Meson exchange models assume that the $\pi\pi$ system is
produced by an intermediate scalar $f_0$ meson 
(often referred to as the ``$\sigma$"), 
and the observed 
$m_{\pi\pi}$-dependence at higher invariant mass 
is due primarily to this meson.
Fits to the dipion decay data 
(excluding the problematical 
$\Upsilon'' \to \pi\pi\Upsilon$)
typically prefer a light, broad state with a
mass near 0.5~GeV;
see for example \cite{Komada:2000ez,Lahde:2001zd,Ishida:2001ve}.

Ideally we would cross an accurate model of the $\pi\pi$ production 
amplitudes into the $\pi$ scattering regime to estimate 
$\pi + (Q\bar Q)$ closed-flavor cross sections, 
but this is not yet possible
because no model gives a good simultaneous description of all the 
experimentally observed $c\bar c$ and $b\bar b$ dipion mass distributions. 
For the present we will simply assume the near-threshold $m_{\pi\pi}$
dependences suggested by the existing models, and evaluate
the cross sections
predicted for $\pi^+ J/\psi \to \pi^+ \psi'$ in each case.

For the gluon radiation and scalar anomaly models we assume 
\be
\langle |{\cal A}(s,t)|^2 \rangle  = c_2 t^2 \ .
\ee  
For a power-law form $\langle |{\cal A}(s,t)|^2 \rangle  = c_n t^n$,
the relation between decays and cross sections Eq.(3)
generalizes as follows; the decay rate Eq.(2) becomes

\be
\Gamma_{{\cal M}_b \to \pi \pi {\cal M}_a }
=
{1\over 256 \pi^3 M_b^3}\; c_n {\cal I}^{(n)}
\ee
where ${\cal I}^{(n)}$ is the integral of $t^n$ over the 
${\cal M}_b \to \pi \pi {\cal M}_a$ Dalitz plot,
\be
{\cal I}^{(n)}
\equiv 
\int \!\!\!
\int \!
ds\, dt \; t^n \ .
\ee
The constant $c_n$ is determined
by the measured dipion partial width using Eq.(6), and
substitution of $c_n t^n$ into the cross section formula
Eq.(1) then gives
\bd
\hskip -3cm \sigma_{\pi {\cal M}_a \to \pi {\cal M}_b } =
\ed
\be
\Gamma_{{\cal M}_b \to \pi \pi {\cal M}_a }\, \cdot
\frac{4\pi^2 M_b^3}{(n+1)\, {\cal I}^{(n)}}\,
\frac{1}{s p_i^{\, 2}} 
\Big(
|t_1|^{n+1} - |t_2|^{n+1} 
\Big) \ .
\ee

We again specialize to the process
$\pi^+ J/\psi \to \pi^+ \psi'$. 
Setting $n=2$ (scalar anomaly models),
for Eq.(7) we find
${\cal I}^{(2)} = 0.01954$~GeV$^8$.
The cross section in Eq.(8) may then be evaluated numerically,
which gives the result shown in Fig.5.
Note that the scalar anomaly model leads to a rapid increase
of the cross section relative to the constant amplitude model 
above $\sqrt{s} \approx 3.9$~GeV. This is due to the 
$t^2$-weighting combined with the rapid 
increase in the range of $t$ covered by this reaction
with increasing 
$\sqrt{s}$, which is evident in Fig.1 (lower right region).

Finally, for our meson exchange model we assume
a generalized Breit-Wigner form which incorporates the scalar anomaly
soft-pion factor. For $t>0$ this is
\be
\langle |{\cal A}(s,t)|^2 \rangle  = 
\frac{\tilde c_2 t^2}{ (\sqrt{t} - M_\s)^2 + \Gamma_{\s}^2/4  } 
\ee  
where $M_\s$ and $\Gamma_\s$ are the 
mass and width of the hypothetical 
scalar meson source of the $\pi\pi$
events. We have incorporated the $t^2$ scalar anomaly soft-pion dependence
in Eq.(9) because pure 
Breit-Wigner forms required 
an unrealistically narrow $f_0$ ($\Gamma_\s \approx 100$-$150$~MeV) and gave
rather poor fits to the data.
In contrast the hybrid form Eq.(9) clearly gives
an acceptable fit (Fig.4), 
although we emphasize that 
{\it this meson exchange model is unphysical}
because the fitted parameters 
$M_\s = 536$~MeV 
and 
$\Gamma_\s = 260$~MeV are inconsistent with the experimental I=0
$\pi\pi$ S-wave phase shift.

We can again use Eqs.(1,2) to determine 
the  
cross section for 
$\pi^+ J/\psi \to \pi^+ \psi{\, '}$ 
implied by this decay model
(ignoring the problem of disagreement with 
phase shifts). The result is 
\bd
\hskip -3cm \sigma_{\pi {\cal M}_a \to \pi {\cal M}_b } =
\ed
\be
\Gamma_{{\cal M}_b \to \pi \pi {\cal M}_a }\, \cdot
\frac{4\pi^2 M_b^3}{{\cal I}_{f_0}}\,
\frac{1}{s p_i^{\, 2}} 
\int_{t_1}^{t_2}  
\frac{t^2 dt}{(\sqrt{-t} + M_\s)^2 + \Gamma_{\s}^2/4} \ . 
\ee
Note that this integration is over negative values of $t$,
whereas the decay rate integral is over 
positive $t$, which leads to different
signs in the Breit-Wigner functions.
The Dalitz plot decay integral 
\be
{\cal I}_{f_0}  
\equiv 
\int \!\!\!
\int \!
ds\, dt \; 
\frac{t^2}{(\sqrt{t} - M_\s)^2 + \Gamma_{\s}^2/4}\ . 
\ee
for $\psi{\, '} \to \pi^+ \pi^- J/\psi\,$
equals $0.08069$~GeV$^6$ 
given our external meson masses and fitted
$f_0$ parameters.
The integral over $t$ in Eq.(10) is
\bd
I = 
M_{\s}^4 
\bigg\{
\frac{1}{2} x^4
- \frac{10}{3}  x^3
+ (10 - c^2) x^2
- (20 -10 c^2) x
\ed
\bd
+( 5 - 10c^2 + c^4)
\ln(x^2 + c^2 )
\ed
\be
-(2 - 20c^2 + 10c^4) 
\frac{\tan^{-1}(x/c)}{c}
\bigg\}
\Bigg|^{x=1+\sqrt{-t_1}/M_{\s}}_{x=1+\sqrt{-t_2}/M_{\s}} 
\ee   
where $c = \Gamma_\s/2M_\s$.
Combining Eqs.(10-12) gives the meson exchange model
prediction for 
$\sigma_{\pi^+ J/\psi \to \pi^+ \psi{\, '}}$, which is   
also shown in Fig.5. Evidently the meson exchange 
cross section is suppressed near threshold,
due to the 
separation from the $f_0$ pole. 

\begin{figure}
\includegraphics[width=0.45\textwidth]{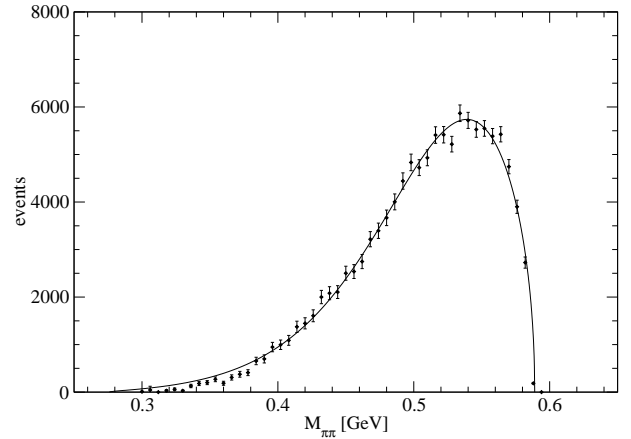}
\caption{
Fitting a meson exchange model to the $\pi\pi$ distribution observed in
$\Gamma_{\psi{\, '} \to  \pi^+ \pi^- J/\psi\, }$.
} 
\label{fig:bk_fig4}
\end{figure}

\begin{figure}
\vskip 1cm
\includegraphics[width=0.45\textwidth]{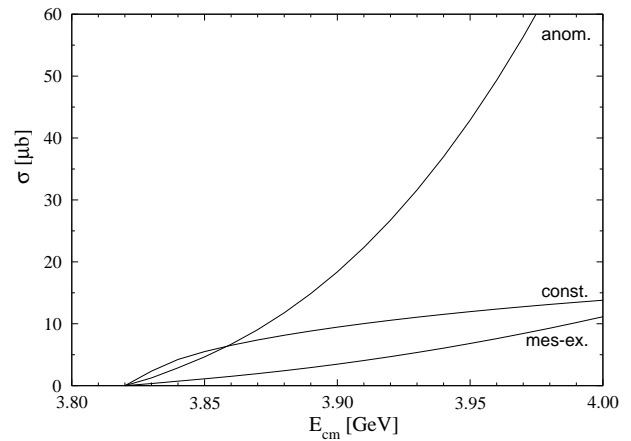}
\caption{
The near-threshold cross section for
$\pi^+ J/\psi \to \pi^+ \psi{\, '}$,  
estimated from 
$\Gamma_{\psi{\, '} \to   \pi^+ \pi^- J/\psi\, }$
in constant amplitude, scalar anomaly and meson exchange
models.
} 
\label{fig:bk_fig5}
\end{figure}
\section{Comparison with previous work}

Two of the theoretical references on dipion decays
cited previously, 
Sorge, Shuryak and Zahed \cite{Sorge:1997bg} and
Chen and Savage \cite{Chen:1997zz}, 
also discussed results for 
closed-flavor pion scattering cross sections with heavy quarkonia, 
specifically for
$\pi^+ J/\psi \to \pi^+ \psi{\, '}$
at low energies. 
Since these references actually
used dipion decay data 
to normalize their scattering amplitudes, approximate agreement
with our results should be anticipated.
Fujii and Kharzeev \cite{Fujii:1999xn} have also evaluated 
closed-flavor charmonium cross sections, using a color-dipole
scattering model. They
do not use dipion decay data as 
a direct input, although they do
note that their two results for the rate 
$\Gamma(\psi{\, '}\to\pi \pi J/\psi)$
are not far from experiment.

In the earliest reference \cite{Sorge:1997bg},
Sorge {\it et al.} assume scalar meson exchange
with a high-mass $f_0(1400)$ as the 
$\pi\pi$ source.
(They make the important and often neglected
observation that assuming a low-mass $f_0$ 
of only moderate width, as in Fig.4,
disagrees with
the experimental I=0 $\pi\pi$ S-wave phase shift.) 
Their cross section close to threshold is shown in Fig.6, and is evidently 
qualitatively similar to our scalar anomaly model result.
Chen and Savage \cite{Chen:1997zz} used a gluon radiation model 
to describe this
reaction, and quoted cross sections 
at tree level and with one-loop chiral
corrections. These results are also shown in Fig.6, and are 
numerically rather similar to Sorge {\it et al}. 
Finally, Fujii and Kharzeev \cite{Fujii:1999xn} 
used a color-dipole scattering model, and quote results for this process
both with and without a scalar $f_0$ form factor for the $\pi\pi$ system.
(Their form factor is inferred from the I=0 $\pi\pi$ S-wave phase shift.) 
Their results without a form factor are similar to the earlier predictions. 
With a form factor they find a much smaller cross section, which  
is rather close to our scalar meson exchange result.

\section{Summary and conclusions}

The search for the quark gluon plasma has led to great interest
in the scale of the cross sections of heavy quarkonia interacting
with light hadrons near threshold. Unfortunately, little is 
known about these cross sections experimentally.
In this paper we have used crossing symmetry and several simple 
amplitude models to estimate the closed-flavor cross sections
for heavy quarkonia scattering on pions
near threshold, using the experimentally
known dipion decays as input. The method is applied both to charmonia
and to the $b\bar b$ system. For the simplest cases of
$1S \lr 2S$ transitions, 
{\it assuming constant amplitudes} 
we estimate the cross sections a few hundred MeV above
threshold to be  
{\it ca.} 20~{$\mu$b}
for 
$\pi J/\psi \to \pi \psi{\, '}$,
and
{\it ca.} 2~{$\mu$b} for
$\pi \Upsilon \to \pi \Upsilon'$. 
The corresponding time-reversed, exothermic 
reactions are estimated to be about 
0.1~{mb} and 10~{$\mu$b} respectively. 

\begin{figure}
\includegraphics[width=0.45\textwidth]{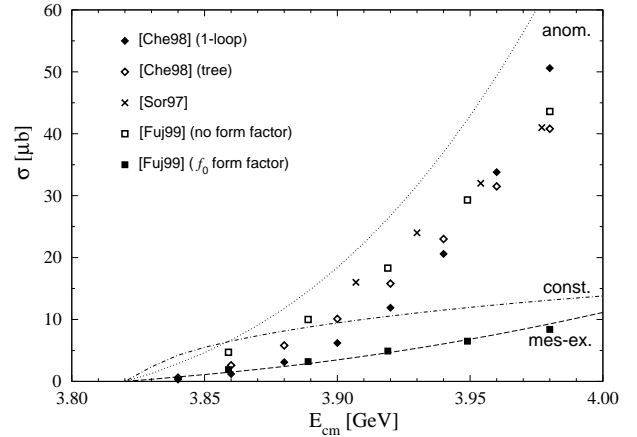}
\caption{
A comparison of our theoretical cross sections for 
$\pi^+ J/\psi \to \pi^+ \psi{\, '}$ (from Fig.5)  
with the results of
Sorge, Shuryak and Zahed \cite{Sorge:1997bg},
Chen and Savage \cite{Chen:1997zz} and Fujii and Kharzeev \cite{Fujii:1999xn}.
Note that of the three forms we have assumed, 
only the lowest cross section (meson exchange model)
gives a good fit to the observed $\pi\pi$ distribution (Fig.4).
} 
\label{fig:bk_fig6}
\end{figure}
We note that the strong dependence of the decay 
amplitude on $t=m_{\pi\pi}^2$ 
observed experimentally in the dipion decays 
near threshold 
makes analytic continuation to 
the pion scattering regime rather problematic. 
The dependence of the cross section 
for $\pi J/\psi \to \pi \psi{\, '}$ 
(used as our example) on the different 
model amplitudes is clearly evident even near threshold 
(Fig.5). For this reaction the various models we considered gave  
consistent 
cross sections of $\sim  5$-$15\, \mu$b at E$_{cm} = 3.9$~GeV, but the
predictions diverged rapidly with increasing invariant mass.

In future, high statistics studies of dipion decays 
at CLEO might provide additional useful information about the
decay amplitude.
In particular, it would be useful to accurately
determine the $s$ dependence of the decay amplitude in the Dalitz plot of 
Fig.1 experimentally, in addition to the already well known
$t = m_{\pi\pi}^2$ dependence.

We note in passing that the cross sections for these 
near-threshold closed-flavor processes 
are much smaller than the millibarn scale typically found for 
open-flavor reactions such as
$\pi J/\psi \to {\rm D}^*\bar {\rm D}$ 
\cite{Martins:1994hd,Wong:1999zb,Barnes:2003dg,
Matinian:1998cb,Haglin:1999xs,Navarra:2001pz,Navarra:2001jy,Duraes:2002ux}. 
The possibility that the rather weak closed-flavor reactions 
are due to open-flavor scattering  
at second order \cite{Moxhay:1988ri,Lipkin:tg,Lipkin:1989tw}, rather than to 
exchange of an unphysically light scalar meson,
is an interesting suggestion which merits future investigation.

\section{Acknowledgements}

We are indebted to 
K.Haglin, M.J.Savage, E.Shuryak, E.S.Swanson and C.Y.Wong 
and our PHENIX colleagues for useful discussions, 
and 
to F.A.Harris for providing the 
$\psi' \to \pi^+\pi^- J/\psi$ data of Bai 
{\it et al.}\cite{Bai:1999mj}, which was used 
to determine the parameters of
the meson exchange model. 
This research was supported in part by
the University of Tennessee, 
the U.S. Department of Energy under contract
DE-AC05-00OR22725 at
Oak Ridge National Laboratory (ORNL),  
and through 
European Union grant INTAS-2000-366 
and Russian Federation grants 
RFBR-01-02-16431, RFBR-02-02-81023 
and RFBR-03-02-17291 
at the Joint Institute for Nuclear Research (JINR).

\vfill\eject

\end{document}